\documentclass[twocolumn,floatfix,altaffilletter,superscriptaddress,preprintnumbers,
               tightenlines,showpacs,showkeys,nofootinbib]{revtex4-1}
\usepackage[utf8]{inputenc}
\usepackage[colorlinks=true,citecolor=blue,linkcolor=blue]{hyperref}
\usepackage[normalem]{ulem}
\usepackage{amsmath,amssymb}
\usepackage{epsfig}
\usepackage{graphicx}               
\usepackage{url}
\usepackage{color}
\usepackage{slashed}
\usepackage{multirow}
\usepackage{placeins}
\usepackage[dvipsnames]{xcolor}
\usepackage{epstopdf}
\usepackage{soul}
\usepackage{tikz}
\usepackage[capitalise, english]{cleveref}
\usepackage{siunitx}
\usepackage{xspace}
\usetikzlibrary{trees}
\usetikzlibrary{decorations.pathmorphing}
\usetikzlibrary{decorations.markings}

\definecolor{jblue}{RGB}{20,50,100}
\definecolor{npurple}{RGB} {153, 51, 204}
\definecolor{wred}{RGB}{217,0,56}
\definecolor{white}{RGB}{255,255,255}

\definecolor{korange}{RGB}{235, 80,  43}
\definecolor{korange2}{RGB}{245, 100,  63}
\definecolor{kyelloworange}{RGB}{255, 210,  110}
\definecolor{kyelloworange2}{RGB}{240, 170,  90}
\definecolor{kred}{RGB}{204,  102, 153}
\definecolor{kpurple}{RGB}{153,  61, 190}
\definecolor{kpurplelight}{RGB}{213,  161, 230}

\DeclareSIUnit\year{yr}
\DeclareSIUnit\pc{pc}
\DeclareSIUnit\ergs{ergs}
\DeclareSIUnit\msun{\ensuremath{M_\odot}}
\allowdisplaybreaks

\makeatletter
\providecommand*{\diff}%
  {\@ifnextchar^{\DIfF}{\DIfF^{}}}
\def\DIfF^#1{%
  \mathop{\mathrm{\mathstrut d}}%
    \nolimits^{#1}\gobblespace}
\def\gobblespace{%
  \futurelet\diffarg\opspace}
\def\opspace{%
  \let\DiffSpace\!%
  \ifx\diffarg(%
    \let\DiffSpace\relax
  \else
    \ifx\diffarg[%
      \let\DiffSpace\relax
    \else
        \ifx\diffarg\{%
        \let\DiffSpace\relax
      \fi\fi\fi\DiffSpace}


\begin{document}


\title{Primordial black holes as a dark matter candidate are severely constrained by the Galactic Center 511 keV gamma-ray line}

\author{Ranjan Laha}
\affiliation{Theoretical Physics Department, CERN, 1211 Geneva 23, Switzerland \\
{\tt  \href{mailto:ranjan.laha@cern.ch}{ranjan.laha@cern.ch}}
{\tt \footnotesize \href{http://orcid.org/0000-0001-7104-5730}{0000-0001-7104-5730} \smallskip}}

\date{\today}

\preprint{CERN-TH-2019-099}

\begin{abstract}
We derive the strongest constraint on the fraction of dark matter that can be composed of low mass primordial black holes by using the observation of the Galactic Center 511 keV gamma-ray line.  Primordial black holes of masses $\lesssim$ 10$^{15}$ kg will evaporate to produce $e^\pm$ pairs.  The positrons will lose energy in the Galactic Center, become non-relativistic, and then annihilate with the ambient electrons.  We derive robust and conservative bounds by assuming that the rate of positron injection via primordial black hole evaporation is less than what is required to explain the SPI/ INTEGRAL observation of the Galactic Center 511 keV gamma-ray line.  Depending on the primordial black hole mass function and other astrophysical uncertainties, these constraints are the most stringent in the literature and show that primordial black holes contribute to less than 1\% of the dark matter density.  Our technique also probes part of the mass range which was completely unconstrained by previous studies.
\end{abstract}

\maketitle

\section{Introduction}
\label{sec:introduction}

Is it possible to constrain the primordial black hole (PBH) density such that it cannot contribute to the entire dark matter density over its viable mass range?  Answering this question will have important implications for the search of the identity of dark matter and inflationary dynamics which can give rise to PBHs\,\cite{1966AZh....43..758Z, Hawking:1971ei, Carr:1974nx, Meszaros:1975ef, Carr:1975qj, Khlopov:2008qy}.  In this Letter, we take one step toward answering this question.  We show that combining the observation that light PBHs can produce $e^\pm$ pairs via evaporation\,\cite{Boudaud:2018hqb} with the fact that an intense 511 keV gamma-ray line has been observed in the Galactic Center\,\cite{Knodlseder:2005yq, Milne:2006ad, Siegert:2019tus, Prantzos:2010wi, Siegert:2015knp, Churazov:2010wy, Jean:2005af, Kierans:2019pkh} can efficiently constrain PBHs in a mass range which cannot yet be constrained by any other technique.  The morphology of the 511 keV gamma-ray line (it has a bulge and a disk component) is such that primordial black holes, acting as the dark matter, cannot explain the entire emission.  We do not yet know the source of these low-energy astrophysical positrons; therefore an understanding of the underlying astrophysical source(s)\,\cite{Lingenfelter:2009kx, Higdon:2007fu, Panther:2018xvc, Alexis:2014rba, Panther:2019fre} can further improve our constraints.

\begin{figure}
\centering
\includegraphics[width=\columnwidth]{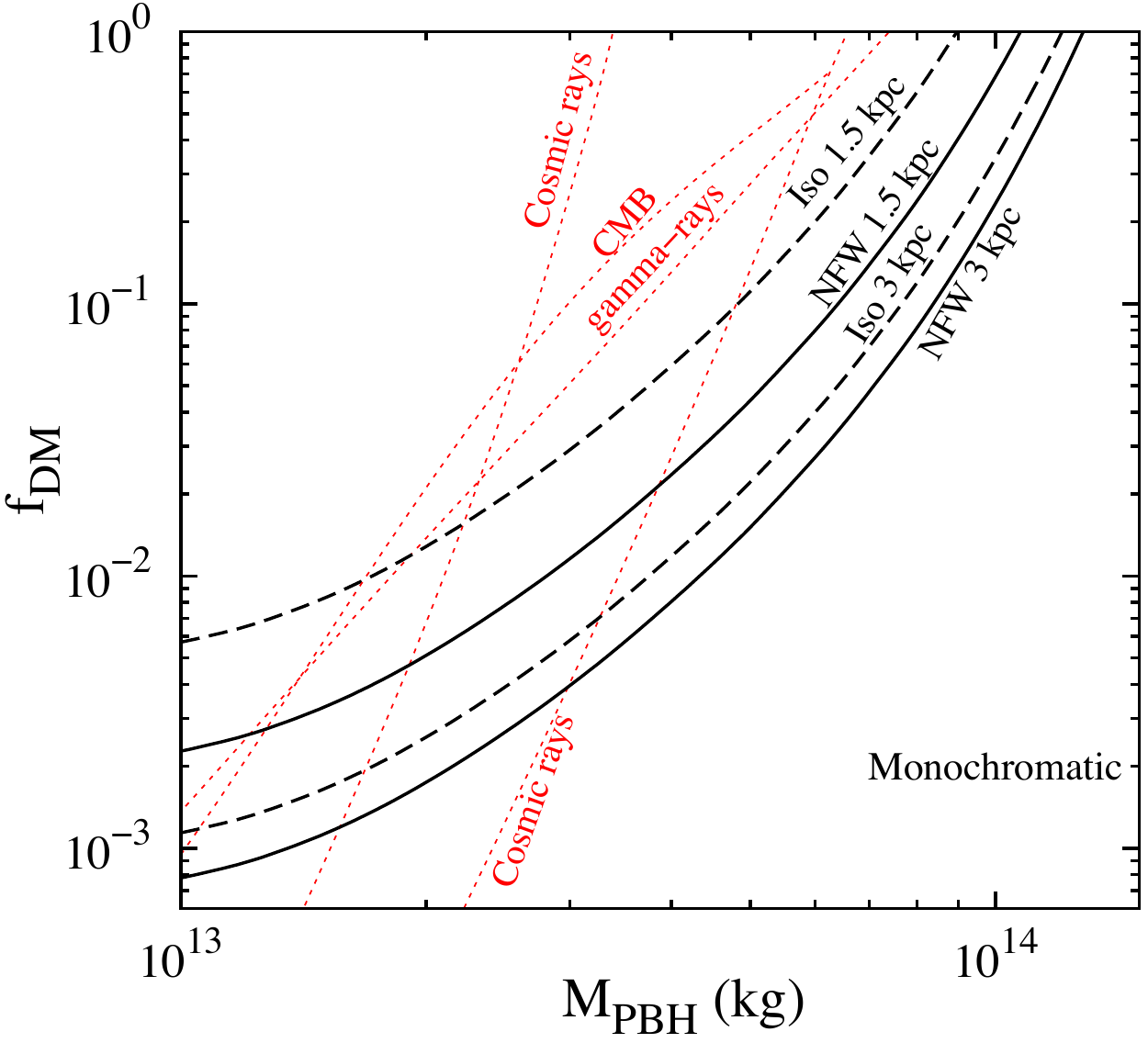}
\caption{Upper limits on the fraction of dark matter which can be composed of primordial black holes for monochromatic mass distribution.  The black lines show the limits derived in this Letter.  These limits depend on the dark matter density near the Galactic Center (NFW and isothermal) and the propagation of low-energy positrons.  While deriving the limit for the ``3 kpc" (``1.5 kpc") constraint, we assume that low-energy positrons can travel about a kpc (100 pc) before annihilating.  The upper limits from Voyager 1 (two lines denote the propagation and background uncertainties)\,\cite{Boudaud:2018hqb}, Planck\,\cite{Clark:2016nst}, and gamma-ray observatories\,\cite{Carr:2016hva, Carr:2009jm} are in red dotted lines.}
  \label{fig:fDM monochromatic}
\end{figure}

The identity of dark matter is one of the most enduring mysteries of physics.  Numerous astrophysical and cosmological observations give an irrefutable indication of the presence of dark matter, yet an absence of its microphysical understanding drives a great deal of research.  A large number of dark matter candidates have been proposed in the literature and these range in masses from $\sim$ 10$^{-22}$ eV to a few hundred $M_\odot$.  PBHs are one of the oldest candidates of dark matter and their abundance has been studied in a number of ways.  The various constraints on PBHs arise from evaporation (and the  subsequent detection of Standard Model particles), capture on astronomical bodies, lensing observations, dynamics of galaxies, gravitational wave observations, and accretion\,\cite{Katz:2018zrn, Graham:2015apa, Montero-Camacho:2019jte, Carr:2009jm, Takhistov:2017nmt, Niikura:2017zjd, Griest:2013esa, Stocker:2018avm, Allsman:2000kg, Tisserand:2006zx, Wyrzykowski:2011tr, Hektor:2018qqw, Clark:2018ghm,  Zumalacarregui:2017qqd, Mediavilla:2017bok, Brandt:2016aco, Koushiappas:2017chw, Monroy-Rodriguez:2014ula, Wang:2016ana, Clesse:2016ajp, Ali-Haimoud:2016mbv, Poulin:2017bwe, Bernal:2017vvn, Sugiyama:2019dgt}.  The recent detection of gravitational waves from binary black hole mergers\,\cite{Abbott:2016blz, LIGOScientific:2018jsj} have rekindled an interest in the contribution of PBHs to the dark matter energy density\,\cite{Bird:2016dcv, Clesse:2016vqa, Sasaki:2016jop}.  This has led to a detailed reanalysis of older constraints\,\cite{Katz:2018zrn, Blum:2016cjs, Ali-Haimoud:2016mbv, Ali-Haimoud:2016mbv, Poulin:2017bwe}, research into new ways to constrain PBHs (for e.g., lensing of fast radio bursts\,\cite{Munoz:2016tmg, Laha:2018zav} and other techniques\,\cite{Jung:2019fcs, Bai:2018bej}), and the study of spinning PBHs\,\cite{Arbey:2019mbc, Arbey:2019jmj, Arbey:2019vqx}.  A detailed study of older constraints has shown that there are viable regions of parameter space where primordial black holes can satisfy the entire dark matter density\,\cite{Katz:2018zrn, Carr:2016drx}.  Ref.\,\cite{Boudaud:2018hqb} pointed out that PBHs with masses $\lesssim$ 10$^{14}$ kg can produce $e^\pm$ pairs via Hawking radiation and such a process can be constrained via the observations of the Voyager 1 satellite\,\cite{2013Sci...341..150S, 2016ApJ...831...18C}.  This naturally leads us to wonder about the fate of $e^\pm$ pairs produced via PBH evaporation in the Galactic Center.  PBHs are much more numerous in the Galactic Center than in the Solar circle (where the Voyager 1 observations were made), and thus a stronger bound can be expected if there is an appropriate observable.  Thus, we are led to the question: is there an observable in the Galactic Center which points to the fact that $e^\pm$ pairs are copiously present there?

The answer is yes.  There is a smoking gun signature which indicates that there is a huge reservoir of low-energy positrons near the Galactic Center.  For many decades, an enduring astrophysical mystery is the observation of 511 keV gamma-ray line at the Galactic Center (see Ref.\,\cite{Prantzos:2010wi} for a historical account). This gamma-ray line has been observed by a number of observatories and a detailed study has been made by the SPI/ INTEGRAL observatory.  Despite the intense scrutiny of this signal, we do not yet know the origin of this signal.  Many viable astrophysical models (i.e., models which do not require a dark matter origin of the 511 keV signal) have been proposed\,\cite{Totani:2006zx, Bandyopadhyay:2008ts, Wang:2005cqa, Milne:2001zs, Bartels:2018eyb, Fuller:2018ttb, Alexis:2014rba, Crocker:2016zzt, Venter:2015gga, Prantzos:2005pz, Weidenspointner:2008zz, Bisnovatyi-Kogan:2016dgr}, although none are confirmed to be the source of these low energy positrons.  A detailed morphological study of this signal and its absence in the dwarf galaxies\,\cite{Siegert:2016ijv} indicate that this it is not produced via dark matter interactions\,\cite{Wilkinson:2016gsy} (see, however, \cite{Farzan:2017hol} for a particle dark matter model which can explain the signal).  Earlier studies trying to connect PBHs and the Galactic Center 511 keV line can be found in Refs.\,\cite{1980AA....81..263O, okeke1980primary, 1991ApJ...371..447M, Bambi:2008kx}.  Thus, any astrophysical source (present in the Galactic Center/ Galactic bulge) which produces low-energy positrons can be constrained via this observation.  The fact that low mass PBHs can produce positrons in large quantities leads us to expect that this observation can be a stringent constraint on the PBH abundance.  

The upper limits on the PBH density assuming a monochromatic mass function derived in this work are shown in Fig.\,\ref{fig:fDM monochromatic} by the black solid and dashed lines.  Various black lines indicate the dependence of this upper limit on the underlying astrophysical parameters.  A part of this parameter space is already probed by gamma-ray observations, cosmic microwave background (CMB) observations, and Voyager 1 measurements.  Our constraints are stronger than these and probe a new mass range of PBHs.  Our constraints close a part of a mass window where PBHs could have contributed to the entire dark matter energy density of the Universe.  Our technique introduces a new electromagnetic probe of PBHs beyond what has already been discussed in the literature\,\cite{Ali-Haimoud:2019khd}.


\section{Formalism}
\label{sec:formalism}

In natural units, the temperature of a black hole of mass $M_{\rm BH}$ is\,\cite{Hawking:1974rv, Hawking:1974sw}
\begin{eqnarray}
T_{\rm BH} = \dfrac{1}{8\pi G_N M_{\rm BH}} = 1.06 \, \left(\dfrac{10^{10} \, {\rm kg}}{M_{\rm BH}} \right) \, {\rm GeV} \, ,
\label{eq:black hole temperature}
\end{eqnarray}
where $G_N$ denotes Newton's gravitational constant.  The temperature of the black hole also dictates the rate at which particles are produced via evaporation.  The energy spectrum of these particles follows the distribution\,\cite{MacGibbon:1990zk}
\begin{eqnarray}
\dfrac{dN_s}{dE} = \dfrac{\Gamma_{\rm s}}{2\pi} \, \int dt \, \dfrac{1}{{\rm exp}(E/T_{\rm BH}) - (-1)^{\rm 2s}} \, ,
\label{eq:evaporation rate}
\end{eqnarray}
where the dimensionless absorption probability is denoted by $\Gamma_{\rm s}$, s denotes the particle spin, and $E$ denotes the energy of the emitted particle\,\cite{MacGibbon:1990zk, Page:1976df, Page:1976ki, Page:1977um}.  Since the positrons in our case if interest are semi-relativistic to non-relativistic, we use the full formula of the dimensionless absorption probability as  in eqn.\,6 of Ref.\,\cite{MacGibbon:1990zk}.  The values of the absorption cross-section $\sigma_s$ is taken from Fig.\,1 of Ref.\,\cite{Page:1977um}.  We also take into account the factor of 2 for the two chiralities of the positron in the full formula of the dimensionless absorption probability.  Since a black hole loses mass via evaporation, $T_{\rm BH}$ is a function of time.  In our calculation, we will use the observed positron injection luminosity of over one year, and the mass loss during this time is negligible for the black hole masses that we consider.  As such, $T_{\rm BH}$ will be a constant for a given black hole mass in our Letter.

The Galactic Center 511 keV gamma-ray line has been observed for a few decades, and its origin has remained unknown throughout.  Recent attempts at measuring the Doppler shifts have also not led to the identification of the source\,\cite{Siegert:2019tus} (note that a similar search technique has also been proposed for the 3.5 keV line\,\cite{Speckhard:2015eva, Powell:2016zbo}).  The observed flux of this gamma-ray line indicates that the rate of positron annihilation at the Galactic Center is $\sim$ 6.3 $\times$ 10$^{50}$ per year\,\cite{Prantzos:2010wi}.  In order to respect the constraints due to the continuum gamma-ray emission measurement, the positrons must be injected at an energy $\lesssim$ 3 MeV\,\cite{Beacom:2005qv}.

\begin{figure*}
\centering
\includegraphics[angle=0.0,width=0.49\textwidth]{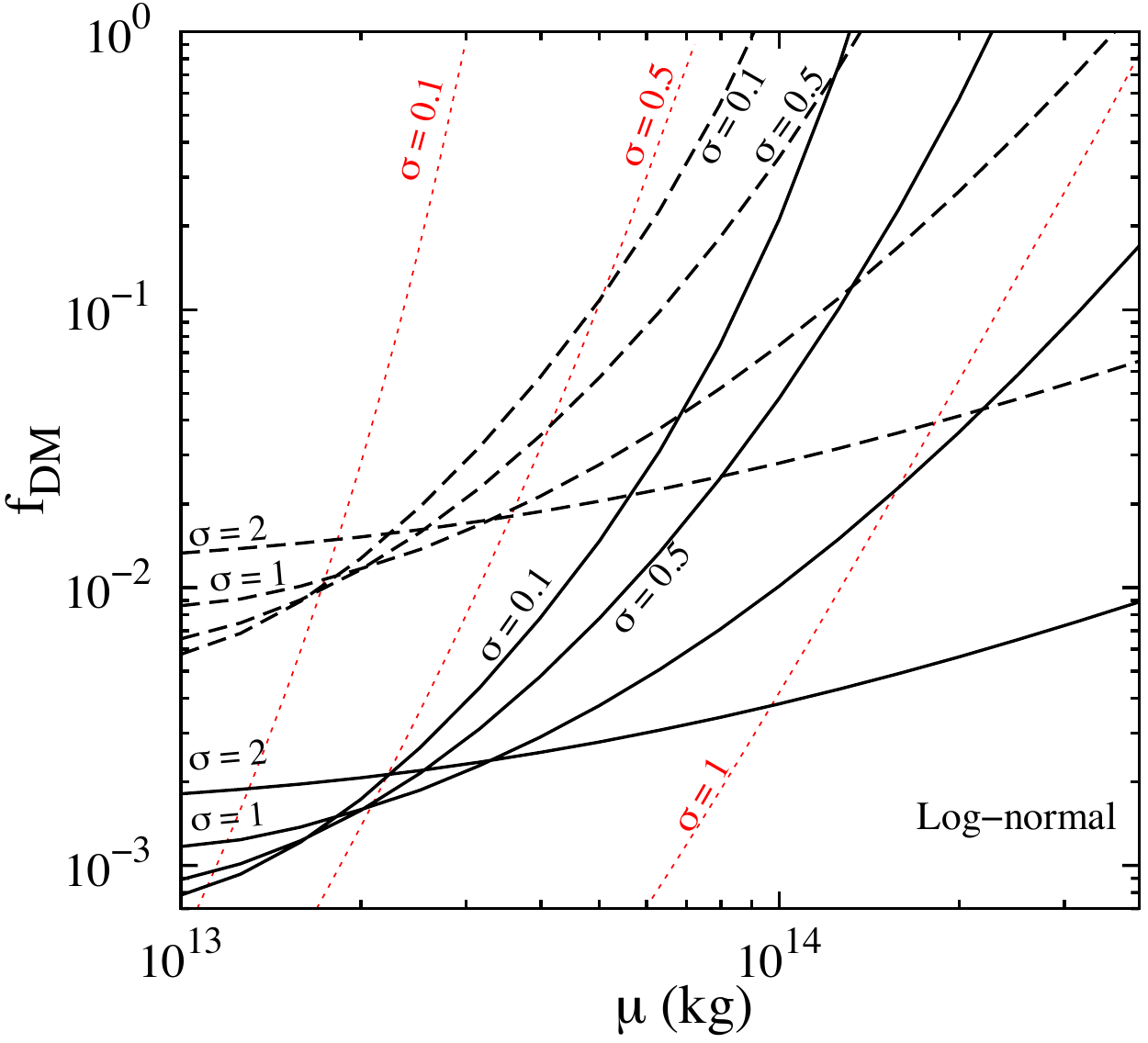}
\includegraphics[angle=0.0,width=0.49\textwidth]{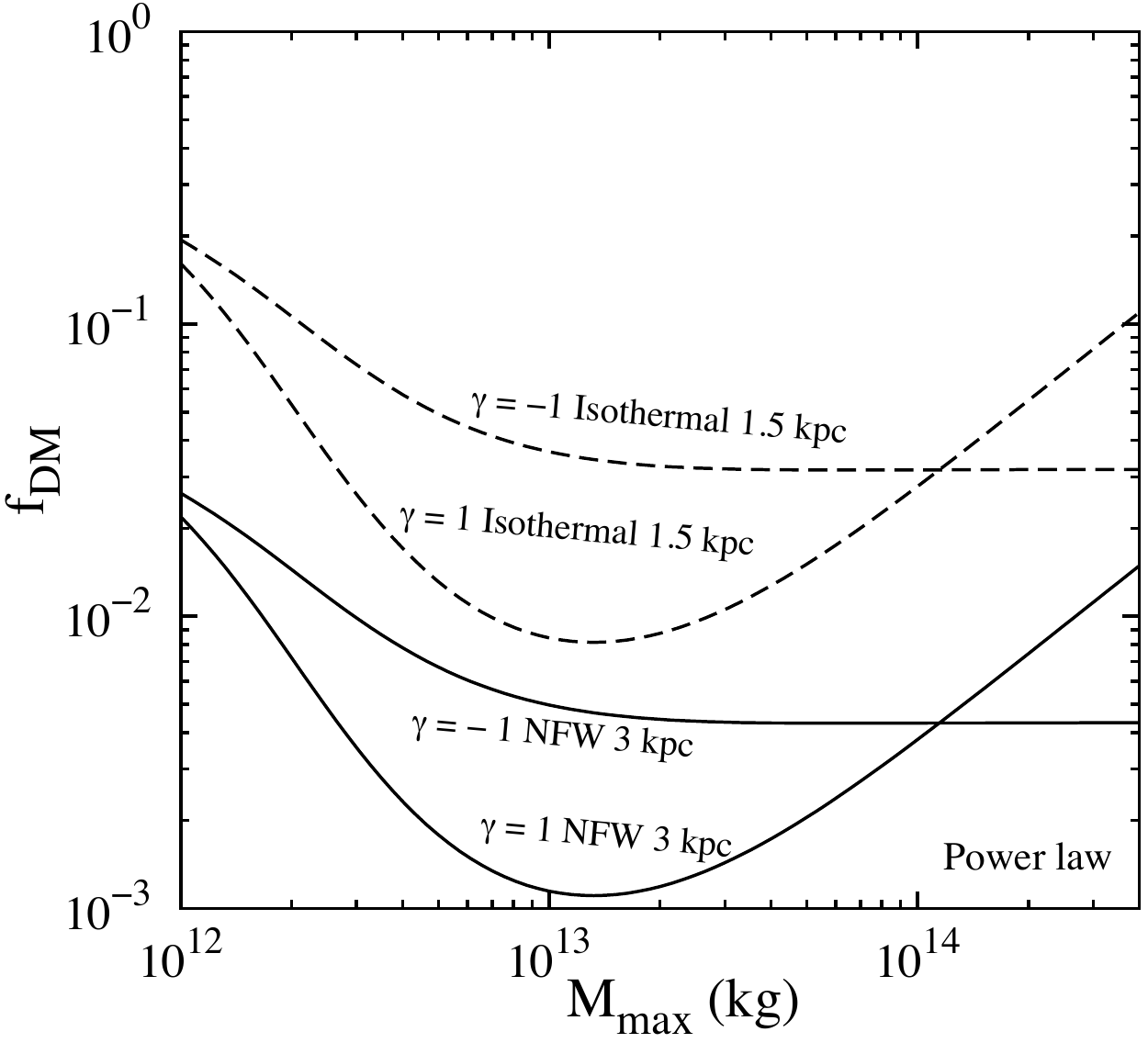}
\caption{Upper limits on the fraction of dark matter which can be composed of primordial black holes for extended mass distribution.  {\bf Left panel:}  We use the log-normal mass distribution with four choices of $\sigma$: 0.1, 0.5, 1, and 2.  The constraints from the Voyager 1 observation is shown as a dotted red line.  When $\sigma$ = 2, Voyager 1 excludes the entire parameter space shown in the figure.  Our limits are shown by the black lines.  The solid (dashed) black line corresponds to the ``NFW 3 kpc" (``Iso 1.5 kpc") assumption.  {\bf Right panel:}  We use the power law mass distribution with two different values of the index, $\gamma$ = $\pm$ 1 and $M_{\rm min}$ = 4 $\times$ 10$^{11}$ kg.}
\label{fig:fDM extended mass distributions}
\vspace{0.1cm}
\end{figure*}

The constraint on the dark matter fraction of PBHs can easily be calculated in the following way.  We first assume that all PBHs have a common mass, i.e., they follow a monochromatic mass distribution.  The number of $e^\pm$ pairs injected per unit time, per unit energy, and per unit volume is given by\,\cite{Boudaud:2018hqb}
\begin{eqnarray}
Q(E, r) = \dfrac{\rho_{\rm DM}(r)}{M_{\rm PBH}} \, \dfrac{d^2N_e}{dE dt} \,,
\label{eq:positrons injected per unit time, energy, and volume}
\end{eqnarray}
where $\rho_{\rm DM} (r)$ denotes the dark matter density as a function of the distance from the Galactic Center, $r$.  The mass of the PBH is denoted by $M_{\rm PBH}$.  In this situation, the upper limit on the fraction of dark matter in the form of PBHs is
\begin{eqnarray}
f_{\rm DM} \lesssim \dfrac{6.3 \times 10^{50}}{\dfrac{\int dV \rho_{\rm DM}(r)}{M_{\rm PBH}} \, \int dE \dfrac{d^2N_e}{dt \, dE} \times 1\, {\rm year}} \, ,
\label{eq:fDM monochromatic upper limit}
\end{eqnarray}
where $dV$ denotes a differential volume element in our Galaxy, and where we use the positron injection luminosity criteria to derive our limits\,\cite{Fuller:2018ttb}.  Note that we simply require that the positron injection luminosity is less than what is measured by SPI/ INTEGRAL.  This is the most conservative option: if we had assumed the contribution of some astrophysical source(s) to this gamma-ray observation, then the upper limit on the positron injection luminosity would have been much lower.  The flux of this line in the bulge and the central source of our Galaxy is (0.96 $\pm$ 0.07) $\times$ 10$^{-3}$ photons cm$^{-2}$ s$^{-1}$ and (0.08 $\pm$ 0.02) $\times$ 10$^{-3}$ photons cm$^{-2}$ s$^{-1}$, respectively\,\cite{Bartels:2018eyb}.  Thus, our limits can be improved by more than an order of magnitude if one identifies the class(es) of astrophysical source(s) which is (are) contributing to this emission in the Galactic bulge.

One source of uncertainty to the upper limit presented in this Letter is the dark matter density profile near the Galactic Center.  We bracket this source of uncertainty by assuming two different dark matter profiles: Navarro-Frenk-White (NFW) and isothermal.  Owing to the cusp present in the NFW profile, the constraints derived using this input is stronger.  We use the forms of the dark matter profiles used in Ref.\,\cite{Ng:2013xha}.  An additional uncertainty arises from the positron propagation distance in the Galactic Center.  The extent of the bulge component of the 511 keV gamma-ray emission is $\lesssim$ 1.5 kpc from the Galactic Center\,\cite{Fuller:2018ttb}.  Non-relativistic positrons can travel a small distance (few hundred parsecs) before annihilation; thus, we assume that all positrons produced from PBHs within 1.5 kpc annihilate in order to derive our upper limits.  However, a clear understanding of positron propagation in the Galactic Center region is still lacking, and there are scenarios via which positrons can be advected up to a distance of $\sim$ 1 kpc\,\cite{Higdon:2007fu, Panther:2018xvc}.  Ref.\,\cite{Higdon:2007fu} finds that 80\% of the positrons born within 3.5 kpc of the Galactic Center annihilate within 1.5 kpc of the Galactic Center.  In order to include this propagation uncertainty, we assume that 80\% of the positrons from PBHs within a 3 kpc radius of the Galactic Center can annihilate within the inner 1.5 kpc.  Our calculations find that the strongest (weakest) upper limits are derived when assuming a NFW (isothermal) profile and including all the PBHs within a 3 kpc (1.5 kpc) radius of the Galactic Center.  These cases bracket the uncertainties due to the dark matter halo and positron propagation.  For all of the cases that we consider, our main conclusions remain true: our limits are more stringent than existing bounds and probe part of the mass range where PBH density was completely unconstrained from previous studies.  The upper limit of the energy integral in Eqn.\,\ref{eq:fDM monochromatic upper limit} is determined by the diffuse gamma-ray data which are produced due to in-flight annihilation of the positrons.  Following Ref.\,\cite{Beacom:2005qv}, we take this upper limit to be 3 MeV.  Since the overwhelming majority of the positrons come to rest before undergoing annihilation, we take the lower limit of the energy integral to be the positron mass.

Various inflationary models also predict an extended mass distribution of PBHs\,\cite{Carr:1975qj, Clesse:2015wea, Carr:2017edp, Hawking:1982ga, Hawking:1987bn, Matsuda:2005ez, Berezin:1982ur, Pi:2017gih, Bhaumik:2019tvl, Escriva:2019phb}.  It is important to study these models in detail as the possibility of whether PBHs contribute significantly to the dark matter density depend substantially on the underlying mass distribution\,\cite{Carr:2016drx, Carr:2017jsz, Green:2016xgy, Kuhnel:2017pwq, Bellomo:2017zsr}.  Traditionally, the limits on the PBH density were presented only for monochromatic mass distributions, however, there has been recent interest in converting these limits for various extended mass distributions of PBHs.  We consider two different classes of extended mass distributions: a log-normal distribution and a power law mass distribution.  A log-normal mass distribution (with $\mu$ and $\sigma$ denoting the mean and the standard deviation of the logarithm of the mass distribution respectively) is defined as
\begin{eqnarray}
\dfrac{d\Phi_{\rm PBH}}{dM_{\rm PBH}} = \dfrac{e^{-\dfrac{\rm ln ^2 \, (M_{PBH}/ \mu)}{2 \sigma^2}}}{\sqrt{2\pi} \sigma M_{\rm PBH}} \, . 
\label{eq:log-normal distribution}
\end{eqnarray}
A power law mass distribution is parametrized by the power law index, $\gamma$, the maximum, $M_{\rm max}$, and the minimum value, $M_{\rm min}$ of the mass distribution:
\begin{eqnarray}
\dfrac{d\Phi_{\rm PBH}}{dM_{\rm PBH}} = \dfrac{\gamma}{M_{\rm max}^\gamma - M_{\rm min}^\gamma} \, \dfrac{1}{M_{\rm PBH}^{1 - \gamma}}\, ,
\label{eq:power law distribution}
\end{eqnarray}
for $M_{\rm PBH}$ $\in$ $[M_{\rm min}, M_{\rm max}]$ and $\gamma$ $\neq$ 0.  In order to derive the limits for these extended mass distributions, we convolve the denominator in Eqn.\,\ref{eq:fDM monochromatic upper limit} with the extended mass distribution:
\begin{eqnarray}
f_{\rm DM} \lesssim \dfrac{6.3 \times 10^{50}/ (1\, {\rm year})}{\int dM_{\rm PBH} \, \dfrac{d\Phi_{\rm PBH}}{dM_{\rm PBH}} \dfrac{\int dV \rho_{\rm DM}(r)}{M_{\rm PBH}} \, \int dE \dfrac{d^2N_e}{dt \, dE}}\,.
\label{eq:fDM extended mass distribution}
\end{eqnarray}

\section{Results}
\label{sec:results}

The results of our Letter is shown in Figs.\,\ref{fig:fDM monochromatic} and \ref{fig:fDM extended mass distributions} for the monochromatic mass function and the extended mass function respectively.  The upper limits for the monochromatic mass function are shown for two different dark matter profiles (the NFW and isothermal profiles) and for two different sizes of the spherical region which hosts the PBHs.  These bracket the uncertainty due to the dark matter profile and positron propagation.  Our constraints are derived by simply assuming that the positrons from PBHs do not overproduce the positron injection luminosity.  As such, our derived constraints are maximally conservative and robust.  

The other constraints shown in the paper are from the observations of Voyager 1\,\cite{Boudaud:2018hqb}, cosmic microwave background\,\cite{Clark:2016nst}, and gamma-rays\,\cite{Carr:2016hva, Carr:2009jm}. The constraints from the Galactic gamma-ray background and the cosmic microwave background is relatively robust with respect to the underlying parameters.  The constraints using the Voyager 1 observations are dependent on the background modeling and cosmic ray propagation uncertainties.  However, even the least stringent bound from Voyager 1 observations is most constraining at low PBH masses.  Depending on the underlying astrophysical parameters, our constraints are the strongest in the mass range $\sim$ 10$^{13}$ kg to $\sim$ 10$^{14}$ kg and probe completely new parameter space.   

We also show our derived upper limits for the log-normal mass distribution of PBHs in the left panel of Fig.\,\ref{fig:fDM extended mass distributions}.  In order to compare our results with the Voyager 1 observations, we use the same values of the log normal mass distribution.  When $\sigma$ = 0.1, our derived constraints are stronger than those derived in Ref.\,\cite{Boudaud:2018hqb} for $\mu$ $\gtrsim$ (1 -- 2) $\times$ 10$^{13}$ kg.  When $\sigma$ = 0.5, our derived constraints are stronger than those derived in Ref.\,\cite{Boudaud:2018hqb} for $\mu$ $\gtrsim$ (2 -- 4) $\times$ 10$^{13}$ kg.  When $\sigma$ = 1, our derived constraints are better than the ones in Ref.\,\cite{Boudaud:2018hqb} only for the NFW 3 kpc case for $\mu$ $\gtrsim$ 10$^{14}$ kg.  The constraints derived in Ref.\,\cite{Boudaud:2018hqb} are better than our constraints in the entire parameter space when $\sigma$ = 2.  This illustrates the complementarity of our constraints with that using the Voyager 1 observations.  Even for the log-normal mass distribution, we find that our technique probes new parts of the parameter space.

Our derived upper limits for the power law distribution is shown in the right panel of Fig.\,\ref{fig:fDM extended mass distributions}.  We assume $\gamma$ = $\pm$ 1 for the two cases that we consider.  We assume $M_{\rm min}$ = 4 $\times$ 10$^{11}$ kg and vary $M_{\rm max}$ to derive our upper limits.  Depending on the dark matter profile and the volume under consideration, the upper limits on $f_{\rm DM}$ are less than 10\% when $M_{\rm max}$ varies between 10$^{12}$ kg to 10$^{14}$ kg.  In this case, we do not show the limits from the Voyager 1 observations as these were not presented in Ref.\,\cite{Boudaud:2018hqb}.

\section{Conclusions}
\label{sec:conclusions}

The direct detection of gravitational waves have started vigorous activities in the field of gravitational wave astroparticle physics.  These observations have prompted many new ideas (see, for example, \cite{Kopp:2018jom}).  This discovery has also led to renewed interest in PBHs as a dark matter candidate.  Remarkably, it has been pointed out by several authors that there are regions of the parameter space where PBHs can contribute to the entire dark matter energy density of the Universe.  In this Letter, we introduce such a technique, which can probe a region of parameter space which cannot be probed by any other observation.  We combine the theoretical insight that low mass PBHs evaporate to produce $e^\pm$ pairs with the SPI/ INTEGRAL observations of the Galactic Center 511 keV gamma-ray line to present the strongest constraint on the PBH density over a wide range in the PBH masses.  We simply require that the positron injection luminosity due to PBH evaporation does not exceed that observed to explain the gamma-ray line observation.  Our constraints depend on the dark matter density and the propagation of low-energy positrons.  Considering all astrophysical uncertainties, we find that our constraints are the most stringent over a wide range in the PBH masses and probe a new part of the parameter space.  An identification of the astrophysical  source(s) responsible for the 511 keV gamma-ray emission can further improve our constraints by more than an order of magnitude.

{\bf Note:}  Ref.\,\cite{PhysRevLett.123.251102}, studying a similar topic, appeared in arXiv while this work was in preparation.  The authors use some different inputs (for e.g., in the upper limit of the volume integration) and our results are comparable.  Both the papers use different inputs and agree on the general message of this technique: this is a new way to probe PBHs in a interesting mass range.

\section*{Acknowledgments}
We thank Jeremy Auffinger, John F.\,Beacom, Mathieu Boudaud, Marco Cirelli, Joachim Kopp, and Toby Opferkuch for discussions and comments.  We especially thank Anupam Ray for extensive discussions.  We thank Jane MacGibbon for pointing out an artefact due to the pdf'ing of her paper and for alerting us to relevant older literature.  We are especially grateful to Don Page for providing us the data of his paper.  R.L. acknowledges the support provided by the CERN Theory group.

\bibliographystyle{JHEP}
\bibliography{refs}

\end{document}